\documentclass[aps,superscriptaddress,twocolumn,showkeys,floatfix]{revtex4-1}

\usepackage{amsmath}
\usepackage{float}
\usepackage{amsfonts}

\usepackage{graphicx}
\usepackage{xspace}
\usepackage{bmpsize}
\usepackage{xcolor}
\usepackage{epstopdf}

\newcommand{\UVfalse}{\ensuremath{(U,V)_{\false}}\xspace}
\newcommand{\UVtrue}{\ensuremath{(U,V)_{\true}}\xspace} 
\newcommand{\true}{\textsc{true}\xspace}
\newcommand{\false}{\textsc{false}\xspace}

\newcommand{\ie}{i.\,e.\xspace}
 
\newcommand{\eg}{e.\,g.\xspace} 

\newcommand{\cf}{cf.\xspace}
\newcommand{\deriv}[1]{\ensuremath{\frac{{\rm d}#1}{{\rm d}t}}}

\begin{document}


\title{ From chemical soup to computing circuit:
  Transforming a contiguous chemical medium into a
  logic gate network by modulating its external conditions}



\author{Matthew Egbert}
\email[]{m.egbert@auckland.ac.nz}
\affiliation{Department of Earth and Planetary Sciences, Harvard University, Cambridge, Massachusetts, USA}
\affiliation{University of Auckland, Auckland, NZ}

\author{Jean-S\'{e}bastien Gagnon}
\email[]{gagnon01@fas.harvard.edu}
\affiliation{Department of Earth and Planetary Sciences, Harvard University, Cambridge, Massachusetts, USA}
\affiliation{Physics Department, Norwich University, Northfield, Vermont, USA}

\author{Juan P\'{e}rez-Mercader}
\email[]{jperezmercader@fas.harvard.edu}
\affiliation{Department of Earth and Planetary Sciences, Harvard University, Cambridge, Massachusetts, USA}
\affiliation{Santa Fe Institute, Santa Fe, New Mexico, USA}


\date{\today}

\begin{abstract}
It has been shown that it is possible to transform a well-stirred chemical medium into a logic-gate simply by varying the chemistry's external conditions (feed rates, lighting conditions, etc).  We extend this work, showing that the same method can be generalized to spatially-extended systems. We vary the external conditions of a well-known chemical medium (a cubic autocatalytic reaction diffusion model), so that different regions of the simulated chemistry are operating under particular conditions at particular times. In so doing, we are able to transform the initially uniform chemistry, not just into a single logic gate, but into a functionally integrated network of diverse logic gates that operate as a basic computational circuit known as a full-adder.
\end{abstract}


\maketitle


\section{Introduction \label{sec:Introduction}}

Digital computers are built from arrays of interconnected logic gates engraved on silicon wafers.  The sheer number of gates that can be placed on a single chip (of the order of a few billions) enables them to process vast quantities of information swiftly and reliably.  Living systems are also capable of processing information (e.g. to transform sensory input into appropriate motor responses \cite{Mitchell_2011,Beer_2014}), and in some organisms, this is accomplished, not through the electrical activity of nervous tissues, but rather by chemistry \cite{Bray_1995}.  The practical implementation of chemical computation is necessarily different from computation accomplished \emph{in silico} (see for example Ref.~\cite{Rozenberg_etal_2012} for a review on natural computing), and it is thus interesting to consider potential new chemistry-based architectures through which living organisms might process information.  A first step toward finding and understanding such architectures is to implement logic gates in a chemical setting.  Many implementations of chemical logic gates are for well-stirred systems where effects due to diffusion can be neglected; see for example the chemical neurons of Refs.~\cite{Hjelmfelt_etal_1991,Hjelmfelt_etal_1992a,Hjelmfelt_etal_1992b,Laplante_etal_1995} and the bistable reaction in a continuously stirred flow reactor of Ref.~\cite{Lebender_Schneider_1994}. In a similar vein, we recently introduced a new method for implementing logic gates in a well-stirred chemical system that involves modifying the chemistry's external conditions  (e.g. feed rates, lighting conditions, etc) in a predefined temporal sequence \cite{egbert_dynamic_2018}.

In order to perform calculations more complex than a single logical operation, it is necessary to connect multiple gates together.  In the laboratory, this has often been accomplished via non-chemical means, \eg by connecting individual gates together with tubes and pumps~\cite{Lebender_Schneider_1994,blittersdorf_chemical_1995} or via electrical coupling.  These approaches require the chemicals comprising each gate to be contained in a vessel, and assume that the chemicals in each vessel are well-stirred.  But most biological systems are extended and not well-stirred.

The goal of this paper is thus to generalize the sequential parametric shifts
method developed in Ref.~\cite{egbert_dynamic_2018} to spatially extended
systems.  We show that with locally modulated external conditions, a contiguous
and initially homogeneous chemistry can be transformed into not just
a single logic gate, but a \emph{network} of interconnected 
logic-gates. In what follows, we explain this method and demonstrate numerically
its application by transforming a spatially-distributed cubic autocatalytic reaction that into a full-adder built from seven logic gates including
NAND, OR and AND.

\section{Theoretical background}
\label{sec:Theory}

We discuss below some features that are necessary to implement a network of logic gates in a spatially extended (i.e. non-well stirred) chemical model based on the scheme developed in Ref.~\cite{egbert_dynamic_2018}.

\subsection{Information propagation in a chemical computer}
\label{sec:Information_propagation}

If a computer is to be constructed from logic gates, it must be possible to assemble the gates into a network where the output of certain gates is used as the input for others.  To date, this has been accomplished in two ways in chemical computing. Some investigations have designed chemical-reaction networks that involve repeated motifs, where each motif corresponds to a logic gate, and additional reactions couple these gates together in order to accomplish the desired computation \cite{moon_genetic_2012,xu_implementation_2013}. For such systems  to function, the reactants of each logic-gate must interact as desired and simultaneously must \emph{not} interact with any of the other logic-gate chemistries, since otherwise undesired `cross-talk' can generate errors and prevent the system from operating as designed. Without suitable spatial compartmentalization, cross-talk and other constraints make it difficult to increase the size and complexity of artificial (and also natural) chemical computational networks. 

A more scalable solution is to do what is done in digital computers: take
advantage of space. In this approach, a more simple chemical system is repeated
in spatial ``tiles''  (where each tile corresponds with a logic
gate---see \eg\cite{lacy_costello_towards_2011}) that are coupled
together by way of their spatial placement. Various mechanisms for transmission
of information between spatially distributed chemical logic gates have been
investigated, such as active transport via pumps~\cite{Lebender_Schneider_1994}
and chemical
waves~\cite{Toth_Showalter_1995,Steinbock_etal_1996,Adamatzky_2004,Costello_Adamatzky_2005,Adamatzky_etal_2005,Adamatzky_Costello_2012,Adamatzky_DurandLose_2012}. 

Perhaps the simplest mechanism of transmission of information between gates would be diffusion, where chemicals in upstream gates passively diffuse into downstream gates. Pure diffusion poses problems however, as a mechanism for information propagation in chemical systems. To be specific, it can be shown that if the upstream and downstream gates are identical (except in their spatial locations), and the only mechanism of transport of information between gates is diffusion, then the state of the downstream gate is a degenerate function of the state of {\em both} gates (a proof of this is provided in Appendix~\ref{appendix:Info_transfer_diffusion}).  This implies that the result of a computation (i.e. final state of the downstream gate) is as much influenced by the information coming from the upstream gate than by the state of the downstream gate leftover from a previous operation, therefore preventing diffusion from being an effective mechanism for information transmission.

In what follows, we show how this issue can be resolved in an initially uniform chemical medium by placing localized regions in different external conditions (or ``parametric regimes'').  This spatial modulation of external conditions imposes a directionality upon the network that allows the information to propagate as desired through the computational network.  Essentially, one spatial region can be placed in a ``receiving'' regime  while its neighbors are in a ``transmitting''  regime, making it possible for diffusion of a single chemical to be the mechanism of information transmission between gates.

\subsection{Modulation of external conditions}

In Ref.~\cite{egbert_dynamic_2018}, we showed how to modulate the feed rates $F_u$ and $F_v$ of the following well-stirred chemical model:
\begin{eqnarray}
\label{eq:Gray_Scott_1}
\deriv{U} &= -r_u U - \lambda U V^2 + F_u(t), \\
\label{eq:Gray_Scott_2}
\deriv{V} &= -r_v V + \lambda U V^2 +F_v(t),
\end{eqnarray}
to configure it to act as a logic gate.  In the above, $U = U(t)$ and $V = V(t)$
are chemical concentrations, $r_{u}$ and $r_{v}$ are decay rates for two
hypothetical chemical species U and V, and $\lambda$ is the rate constant for
the autocatalytic reaction U$+$2V $\rightarrow$ 3V.  Note that the
method of modulation of external conditions presented in
Ref.~\cite{egbert_dynamic_2018} is very general and could in principle be
applied to many systems (including chemical systems).  It is thus difficult to
present a general way of choosing the parameters to be modulated, but two
criteria are important.  The first one is that the parameter must be easily
varied externally by some outside entity.  The second criterion is that the
parameter must sufficiently influence the system such that it enables toggling
between different regimes (e.g. bistable and monostable, see below).  As
presented in \citep{egbert_dynamic_2018}, the inflow of chemicals U and V into
the system ($F_{u}$ and $F_{v}$) satisfy both criteria. 

When $F_u=20$ and $F_v=0$ (with $r_{u} = 1.5$, $r_{v} = 3$ and $\lambda = 1$) this system is bistable, and we associated the values of $U$ and $V$ at the two fixed points with the Boolean values \true and \false thus: $\UVfalse\approx(12.7,0)$ and $\UVtrue\approx(0.5,6.5)$ respectively.  We then showed how the feed rates can be modulated according to a predetermined temporal sequence in a way that causes the system to operate as a logical NAND gate, \ie transforms the initial conditions that correspond to the four possible \true/\false input combinations into the states associated with a NAND operation.  The above procedure enables one to construct one logic gate in a well-stirred system (such as a beaker).  Many such systems could in principle be connected together (via tubes for example) to form a logic gate network, although the final contraption is not easily scaled (as argued in Sect.~\ref{sec:Information_propagation}).  

In this paper, we extend the system spatially and include the effects of diffusion in the chemical kinetics.  Equations~(\ref{eq:Gray_Scott_1})-(\ref{eq:Gray_Scott_2}) thus become:
\begin{eqnarray}
\frac{\partial U}{\partial t} & = & D_u \nabla^2 U -r_u U - \lambda U V^2 +  F_u(x,t) \label{eq:rd_dudt}, \\
\frac{\partial V}{\partial t} & = & D_v \nabla^2 V -r_v V + \lambda U V^2 +  F_v(x,t) \label{eq:rd_dvdt},
\end{eqnarray}
where $D_{u}$ and $D_{v}$ are the diffusion coefficients of chemical species U and V, respectively.  
Below, we show that using a similar approach of varying feed rates as introduced in Ref.~\cite{egbert_dynamic_2018}, it is possible to instantiate multiple chemical logic gates and to compose them into a network, where the output of gates is used as the input of other gates and the mechanism of information transmission between gates is diffusion. The primary difference between the method employed here and that used in Ref.~\cite{egbert_dynamic_2018}, is that in addition to modulating feed rates as a function of time, different spatially localized regions are driven with different feed rates, \ie the parameters can be varied as a function of time \emph{and space}. This allows us to produce concurrently operating spatially-distributed gates that can exchange chemical information in a directed manner by way of diffusion, and overcome the degeneracy argument raised above.

The next section explains the model and how its parameters are modulated as a function of space and time to produce a full-adder (for an introduction to such circuits, see Ref.~\cite{Crowe_Hayes_1998}).


\begin{figure*}[h]
\centering
\includegraphics[width=0.9\linewidth]{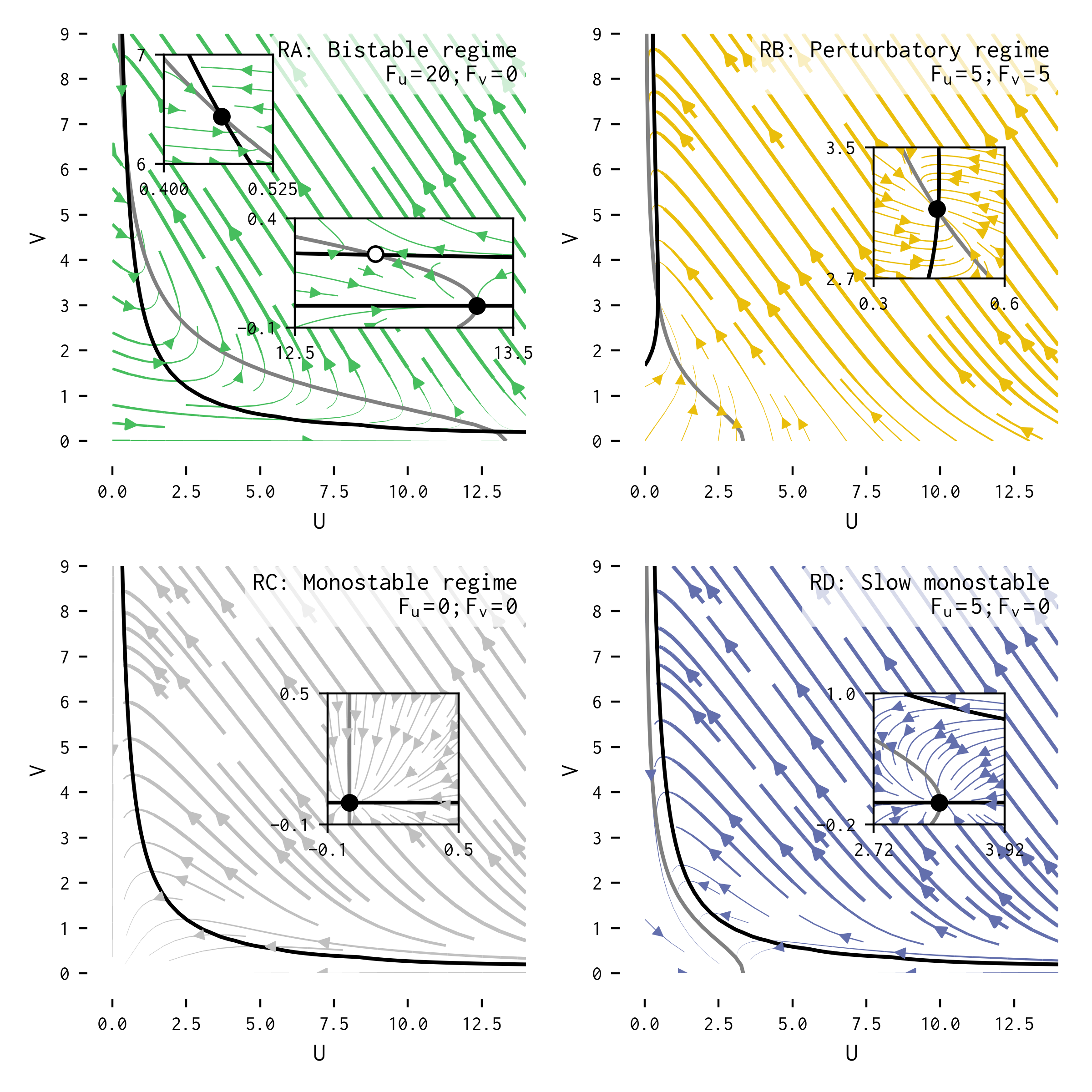}
\caption{Dynamics for the four different parameter regimes (shown in Table~\ref{tab:regime_parameter_values}) of the chemical system represented by Eqs.~(\ref{eq:Gray_Scott_1})-(\ref{eq:Gray_Scott_2}).}
\label{fig:regimes}
\end{figure*}

\section{Method and Results}
\label{sec:Results}

To numerically integrate the reaction-diffusion system defined by
Eqs.~(\ref{eq:rd_dudt})-(\ref{eq:rd_dvdt}), we use the Euler method with
timestep $\Delta t = 10^{-3}$ and spatial resolution $\Delta X = 0.25$.
 In what follows, the time variable (t) and spatial variables (x and y) are expressed in multiples of the timestep and spatial resolution, such that, e.g. 1 time unit corresponds to 1000 integration time steps.   Without
loss of generality, we choose the values $D_{u}=D_{v}=1$, $r_{u} = 1.5$, $r_{v}
= 3$ and $\lambda = 1$ for the purpose of numerical simulations; other values
would change the details of our results, but not the overall reasoning.  
The simulated space consists of `walls,' and `chemistry'. In the latter case, $U(x,t)$ and $V(x,t)$ are dynamic variables representing the concentration of the two reactants. There is no interaction or flux between walls and chemistry, but wherever the chemistry region solution is contiguous, diffusion operates uniformly.  The chemistry area is subdivided into regions. There is no physical boundary between these regions (\ie diffusion occurs uniformly throughout the chemistry), but the feed rate parameters $F_u$ and $F_v$ can be different for different regions.

\begin{table}[tbhp]
\centering
\caption{Parameter regimes}
\begin{tabular}[c]{lcc}
\bfseries Regime & ${\bf F_u}$ & $\bf F_v$ \\
RA (bistable)        & 20 & 0 \\
RB (perturbatory)         & 5  & 5 \\
RC (monostable)     & 0  & 0 \\
RD (slow monostable) & 5  & 0
\end{tabular}
\label{tab:regime_parameter_values}
\end{table}

Each region corresponds to a single logic gate. We use three different types of
logic gate: NAND, OR and AND. For each of these gates, there is a single
``regime sequence'', corresponding to a fixed and periodic temporal sequence of
the regimes shown in
Table~\ref{tab:regime_parameter_values} (for an animated plot showing the working of a single gate, see movie S1 in the Supplementary Information~\cite{Note2}).  The parameter regimes employed here are the same as in Ref.~\cite{egbert_dynamic_2018}, but with an additional fourth regime, referred to as `RD', or the `slow monostable' regime (see Table~\ref{tab:regime_parameter_values} and Fig.~\ref{fig:regimes}). The duration of each regime in a particular regime sequence depends on the gate.  There are likely many possible sequences capable of causing  the chemical system to act as NAND (or other) logic gates. To develop the sequences presented here, we started from the non-spatial model described in Ref.~\cite{egbert_dynamic_2018}, and used simulation and analysis to identify the dynamics required to produce composable logic networks, and then experimented with different regimes and regime sequences to identify how to modulate the system to perform as desired.

\subsection{A reaction-diffusion NAND gate}

\begin{figure*}[ht]
  \centering
  \begin{tabular}{r}
    \includegraphics[width=1.0\linewidth]{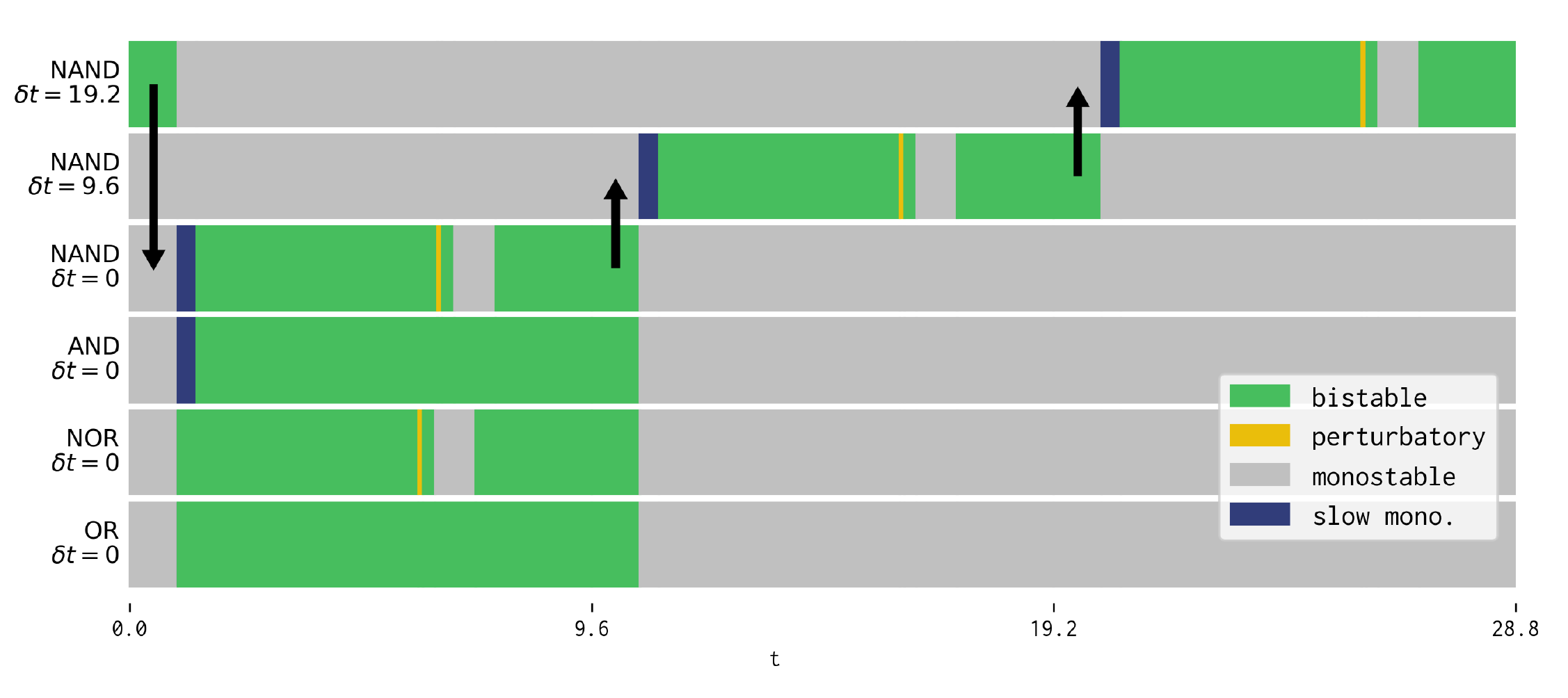}\\ 
  \end{tabular}
  \caption{Regime sequences for time-shifted variations of the NAND regime-sequence (top) and for the other three logic gates (bottom). Transmission between adjacent gates occurs at the times indicated by the arrows, such that the final state of NAND is timed to act as the input for gates with a time shift $\delta t = 9.6$, which in turn can act as input for gates with a time shift of $\delta t = 19.2$, which can act as input for gates with a time shift of $\delta t = 0$.  \label{fig:regime_sequences}}
\end{figure*}

The regime sequence that we derived for spatially distributed, composable NAND gates is similar to that described for the well-stirred system presented in Ref.~\cite{egbert_dynamic_2018}, but now includes additional regimes that we found to be necessary for the diffusion-based inter-gate transmission of information.  As mentioned above and detailed in Appendix~\ref{appendix:Info_transfer_diffusion}, if an upstream and downstream gate are identical and synchronized, it is impossible to distinguish between the influence of the upstream gate and the influence of the downstream gate's previous state. We resolve this by resetting the downstream gates using a `base' regime (corresponding to regime RC in Table~\ref{tab:regime_parameter_values}) before placing them in a `receiving' regime, that is coordinated to be simultaneous with the `transmitting' regime of the upstream gates. This approach requires neighbouring regions to be placed in different regimes even if they are the same type of gate. The simplest way that we found to do so was to use a single repeating regime-sequence for each type of gate and to offset the starting time of these regimes by an amount $\delta t$ so that the transmitting regime of upstream gates coincides with the receiving regime of downstream gates~\cite{Note3}.  Fig.~\ref{fig:regime_sequences} shows the regime sequences for each of the gates and how the time-shifted regimes for the NAND gates line up, so that information can propagate between gates (black arrows in figure).  Note that a minimum of three time-shifted zones are necessary for a network of gates to work. Time-shifted zones enable the directed transmission of information by diffusion such that when a gate is transmitting, its upstream neighbour is {\em not} receiving, but its downstream neighbour is.  For instance, in a network of three gates A, B, C (where B is downstream from A and C is downstream from B) with only two time-shifted zones, A and C would be in the same regime when B was transmitting, and thus information would ``go backward'' in the network, thus preventing effective computation.

\begin{figure}[tbhp]
\centering
  \begin{tabular}{r}
    \includegraphics[width=0.93\linewidth]{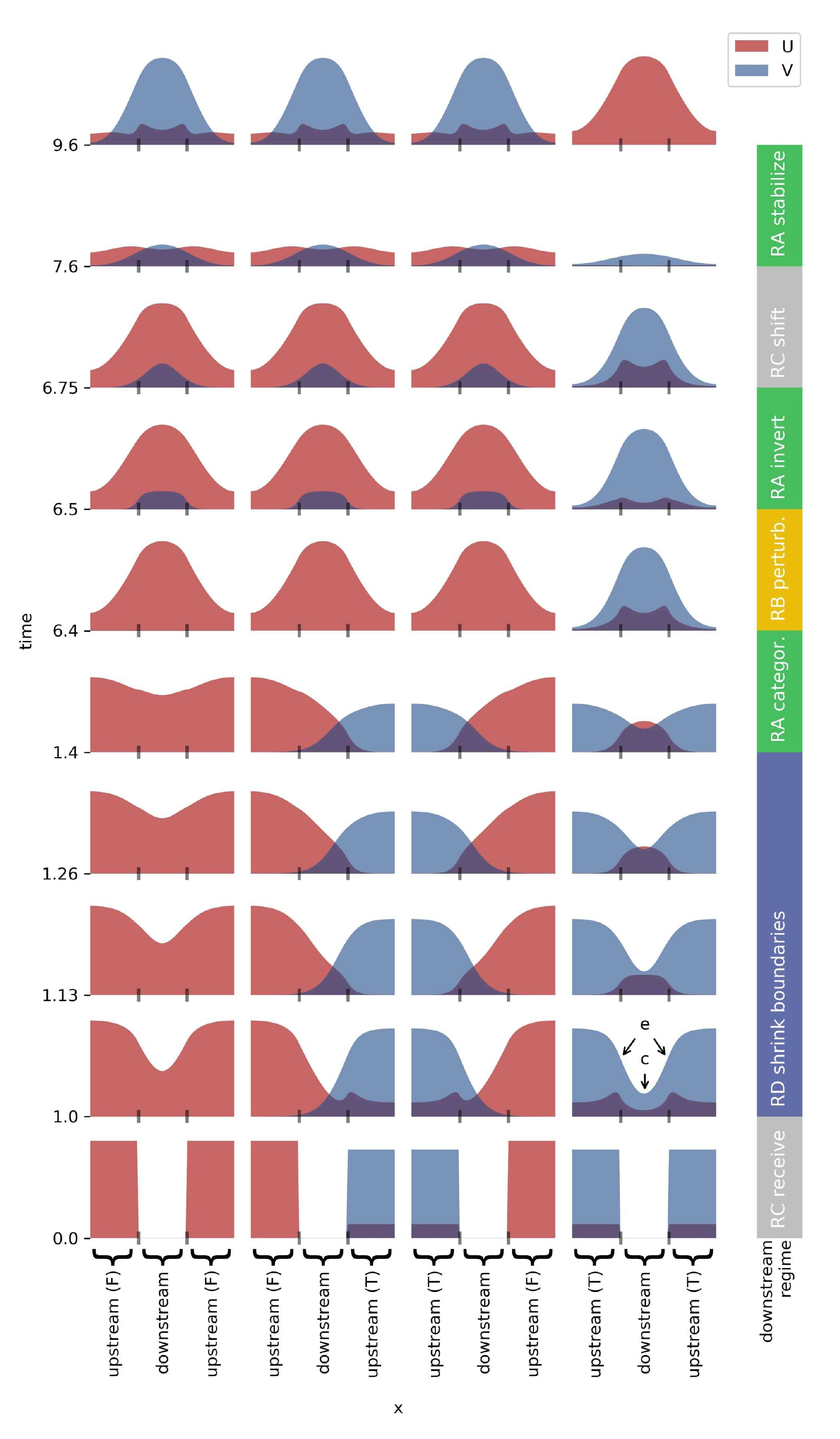}\\ 
  \end{tabular}
  \caption{One-dimensional reaction-diffusion NAND gates. Each column indicates
    a downstream NAND gate, flanked by its two upstream gates.  
    Small black ticks indicate the boundary between the gates, showing the
    regions that are placed in different regimes. The stationary concentrations at $t = 9.6$ correspond to the fixed points of the bistable regime in Fig.~\ref{fig:regimes} (upper left panel).  The gates work
    effectively with widths between $2.0$ and $3.4$ as described in
    Appendix~\ref{appendix:Details_NAND_operation}. Diffusion
    operates uniformly, \ie there is no barricade or other impedance of
    diffusion between gates. The time of each snapshot is indicated on the left
    of the figure, and the columns on the right indicate which regime the
    downstream and upstream gates are in as time progresses from the bottom of
    the figure to the top.  The labels `e’ and `c’ are used in Appendix~\ref{appendix:Details_NAND_operation} where we describe in detail the operation of this gate.  A movie of this process is provided in the
    Supplementary Information (Movie S2).  \label{fig:cross_section_NAND}}
\end{figure}

The logical operation proceeds in three broad steps, involving the receipt of information (i.e. ``high'' or ``low'' chemical concentration of species V) from upstream gates (Steps 1,2 \& 8 in Table \ref{tab:NAND_regime_sequence_specification}), the categorization of that input by target output value---i.e. an error correction step (Step 3), and the transformation of those categorized values into the correct logical value (Steps 4--7). A summary of those steps is provided in Table~\ref{tab:NAND_regime_sequence_specification} and illustrated in Figs.~\ref{fig:regime_sequences} and~\ref{fig:cross_section_NAND}.  The details of the steps for the NAND operation and how the dynamic modulation of the chemistry's operating conditions is used to accomplish them are intricate, and so we have included a detailed description in Appendix~\ref{appendix:Details_NAND_operation}. Again the steps are similar to those described in Ref.~\cite{egbert_dynamic_2018}, but include a few additional steps to facilitate the transmission of information between gates.

NAND gates are universal~\cite{Crowe_Hayes_1998}, so strictly speaking it is unnecessary to implement any other gates directly, but having other gates available generally reduces the total number of gates necessary for implementing a particular operation. Once we identified a sequence of regimes for producing the NAND operation, it was comparably easy to identify related sequences that operate as other gates, including AND, NOR and OR. The details of these gates are outlined in Appendix~\ref{appendix:Other_gates}.


\begin{table}[tbhp]
\centering
\caption{Regime sequence for spatial NAND gates.}
\begin{tabular}[c]{rllc}
    \#  & \bfseries Step Name & \bfseries Regime  & \bfseries Duration \\ 
    1 & Receive               &  RC (monostable) & 1.0 \\
    2 & Shrink boundaries     &  RD (slow monostable)  & 0.4 \\
    3 & Categorize            &  RA (bistable) & 5.0 \\
    4 & Perturb               &  RB (perturbatory) & 0.1\\
    5 & Invert                &  RA (bistable) & 0.25\\
    6 & Shift to separatrix   &  RC (monostable) & 0.85\\
    7 & Stabilize             &  RA (bistable) & 2.0 \\
    8 & Transmit              &  RA (bistable) & 1.0 \\
    9 & Reset                 &  RC (monostable) & 18.2
\end{tabular}
\label{tab:NAND_regime_sequence_specification}
\end{table}

\subsection{Transforming a homogenous chemistry into a 7-gate full-adder}

\begin{figure}[t]
  \centering
  \begin{tabular}{r}
    \includegraphics[width=1.0\linewidth]{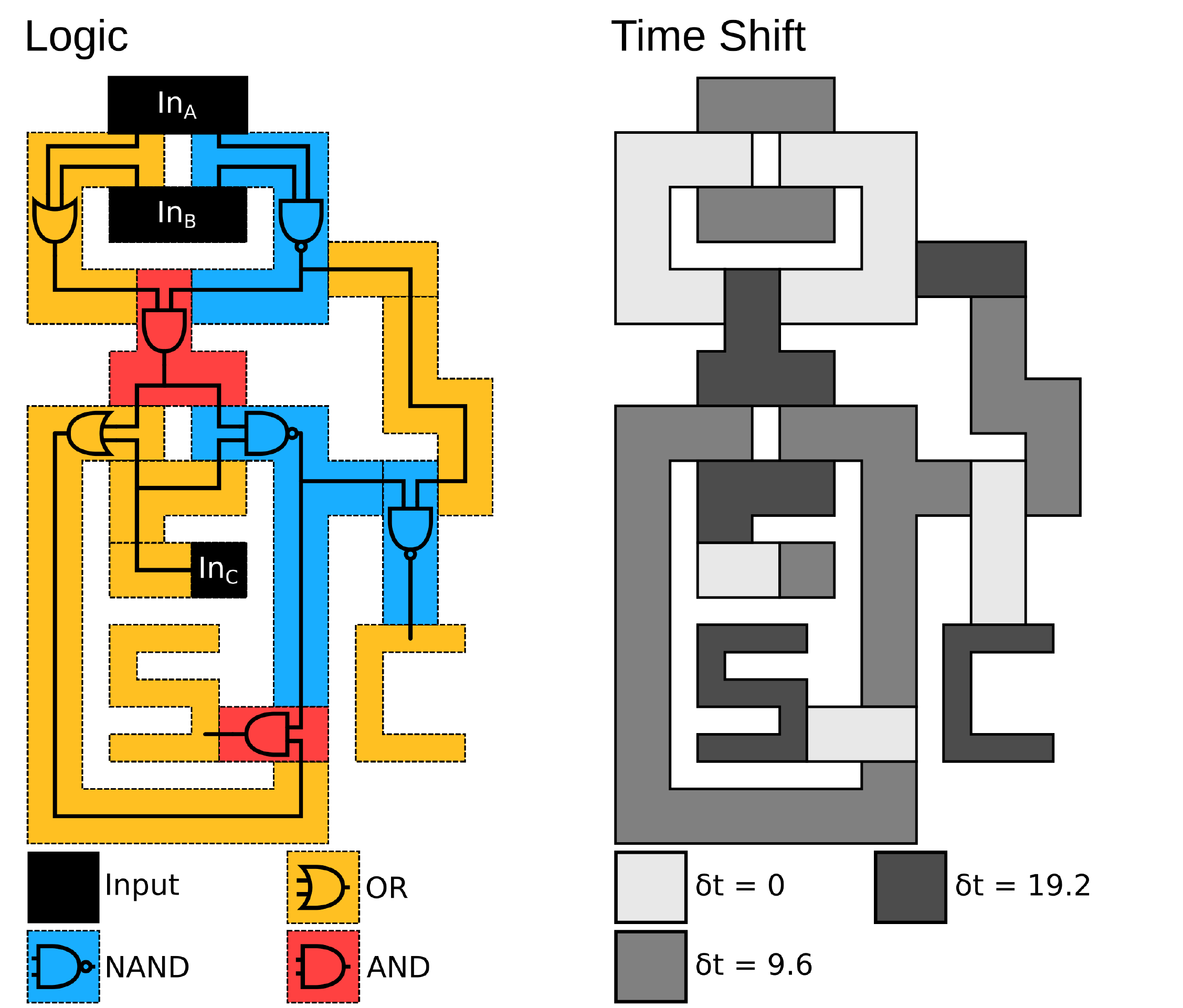}\\ 
  \end{tabular}
  \caption{Full-adder system, showing arrangement of regime-sequences (left) and time shifts (right).  The outputs of the full-adder (sum and carry) are indicated by the ``S'' and ``C'' symbols, respectively.  The gates work
    effectively with widths between $2.0$ and $3.4$ as described in Appendix~\ref{appendix:Details_NAND_operation}  (figure to scale) 
    \label{fig:fulladder}}
\end{figure}

To evaluate the effectiveness of these gates, and to ensure that they can be composed into a logic gate network, we implemented a full-adder. A full-adder adds together two binary digits, calculating the sum and the carry value for the addition. The design for this system is shown in Fig. \ref{fig:fulladder}. In this figure, the white area is a wall where no chemicals exist and no diffusion through the walls takes place. The entire colored area represents a single contiguous reservoir in which different regions are driven by different feed rates. Note that there is no physical boundary between these regions. How the feed rates are modulated is indicated by the two plots in the figure, with the colored plot on the left indicating which sequence (\cf Table \ref{tab:NAND_regime_sequence_specification}) is used for which region, and the values in the right plot indicate the time shift of the regime sequence for each region.

In the full-adder network, the inputs are specified as initial conditions for the three black locations. The network is simulated for five full sequences ($5\cdot 28.8 = 144$ time units), thus allowing for five sequential logic-operations. At the end of this time, the ``sum'' output is indicated by the state of the region shaped as an `S', and the ``carry'' output is indicated by the state of the region that is shaped as a `C'. Note also that we use the OR gate sequence as information conducting `wires' that propagate unmodified Boolean values when only coupled to a single upstream gate (see \eg the rightmost region in Fig.~\ref{fig:fulladder}).
The network produces correct output for all possible input configurations as defined by the full-adder truth table (see Table \ref{tab:full_adder_truth_table}).  A video of a test of six examples with different inputs is included in the Supplementary Information (Movie S3).

\begin{table}[tbhp]
\centering
\caption{Full-adder truth table.}
\begin{tabular}[c]{ccc|cc}
    Input A &  Input B & Input C & Sum & Carry \\ \hline
    \false & \false & \false & \false & \false \\
    \false & \false & \true  & \true & \false \\
    \false & \true & \false  & \true & \false \\
    \false & \true & \true  & \false & \false \\
    \true & \false & \false  & \true & \false \\
    \true & \false & \true  & \false & \true \\
    \true & \true & \false  & \false & \true \\
    \true & \true & \true  & \true & \true
\end{tabular}
\label{tab:full_adder_truth_table}
\end{table}

\section{Discussion}
\label{sec:discussion}

The method developed in this paper is a spatial extension of that presented in Ref.~\cite{egbert_dynamic_2018}.  The fundamental idea common to both methods is the use of sequential shifts of the external environment to transform the dynamics of a chemical system.  In this paper, we have considered a chemical system that is spatially extended and not well-mixed. In this more general context, we have shown that it is possible to transform a simple non-linear chemical soup, not just into a logic-gate, but a network of such elements. In addition to enabling the instantiation of numerous gates within a single contiguous medium, the spatiotemporal modulation of environmental conditions enables the imposition of asymmetrical conditions upon the medium that allows the use of diffusion as the mechanism of inter-gate information propagation.  Note that in our system, the information propagates between gates (i.e. from an upstream gate to its downstream neighbour) by way of diffusion, but the output state produced by the logical operation propagates from its input region to its output(s) by way of a reaction-diffusion travelling wave that spreads across the gate region (see Supplementary Information Movie S3). The primary advantage of diffusion-based inter-gate interaction is that it allows everything to be accomplished in chemistry, rather than requiring separate mechanisms for transmitting state between logic gates.  We demonstrated our new method by transforming a spatially distributed cubic-autocatalytic reaction into seven logic-gates that are connected so as to implement a binary addition circuit known as a full-adder.



As discussed in the introduction, previous work in the area of chemical computing has used either spatial compartmentalization or artificial chemical-network construction to create chemistry-based logic networks; methods that have proven difficult to scale up beyond a small number of logic-gates. The method introduced here may indicate an alternative, more scalable route for creating chemistry-based logic-gate networks and computers.

In addition, since the programming is done externally by modulating (both in time and space) the operating conditions of the chemical system, it may also be possible to reprogram the same medium in real time, i.e.~during or in between computations.  
This reprogrammability feature represents one advantage of this type of chemical
computing over conventional digital computers.

The method presented in this paper is based on the sequential parametric shift method of Ref.~\cite{egbert_dynamic_2018}, and is thus subjected to the same constraints.  In particular, the dynamics of parametric changes has to be fast compared to the dynamics related to Eqs.~(\ref{eq:rd_dudt})-(\ref{eq:rd_dvdt}).  If this is not the case, additional equations (coupled to Eqs.~(\ref{eq:rd_dudt})-(\ref{eq:rd_dvdt})) have to be considered to properly take this parametric change transient dynamics into account.  Similarly, we do not know of a general method for identifying a sequence of
external conditions that transforms a chemical or other medium into a
computational device (this is true for the method presented in here and the one in Ref.~\cite{egbert_dynamic_2018}). But through trial and error and exploratory modeling, we have demonstrated that it is possible to do so in some circumstances. In our
investigation we identified a few key conditions that we believe are necessary,
but not always sufficient for accomplishing this kind of manipulation. Specifically, to create binary logic, the underlying medium must include a regime with bistability (\ie, a regime in which there exist at least two stable attractors. Without this property, it would be difficult to avoid the accumulation of any error or noise that imposes itself on the network. In our system, the categorization step accomplished a kind of `error correction' where small fluctuations in state are lost as any trajectories approach the equilibria in Regime-A. In addition, an essential property for universal binary computation is the ability to ‘invert’ truth values, i.e. to either be able to transform an input of.  In our system, we accomplished this be piecing together transients that essentially swap the U and V concentrations of the trajectories (Steps
5--7), i.e. high U, low V are transformed into high V, low U states and vice versa. Without such a step, AND and OR gates are possible, but these are not universal (i.e. unlike NAND and NOR gates, these cannot be combined to create any other binary logic gate).

To our knowledge, the method presented here is significantly different from others in the literature.  Most theoretical and experimental implementations of logic gates in spatially extended media rely upon propagation of chemical waves.  For instance, the logic gate design of Ref.~\cite{Toth_Showalter_1995} involves sending chemical waves into capillary tubes, and reading the output at some crossing point.  Similarly, the work in Ref.~\cite{Steinbock_etal_1996} involves the propagation of chemical waves on a ``printed circuit'' of Belousov-Zhabotinsky catalyst instead of capillary tubes, while the architectureless collision-based computing of Refs.~\cite{Adamatzky_2004,Costello_Adamatzky_2005} relies on transmitting signals via compact excitations in the medium.


The method presented in Ref.~\cite{egbert_dynamic_2018} provides a new way to transform a well-stirred chemistry into a logic-gate (or other `information-processing' operator) by modulating the external conditions (feed rates, lighting conditions, etc) in which the chemistry operates.  We also argue in Ref.~\cite{egbert_dynamic_2018} that this method might be a potentially useful way for living systems to process information, since it does not require specific `chemical hardware' to work, but relies instead on re-purposing existing chemical hardware by modulating the  conditions in which that chemistry operates.  The same can be said of the method presented here.  We have shown that when these conditions are varied in a localized manner, with different regions of the same basic chemistry operating in different parameter regimes, it is possible to create not only a logical operation, but a network of logical operations.

On a more speculative note, it is interesting to consider how biology might similarly take advantage of this kind of architecture. Evolution has no requirement that its product be simple or understandable, only that it works, and so it seems unlikely that it would produce something exactly like what we have engineered here. However, living systems may well take advantage of the changes induced in chemical dynamics by the environment in which the chemistry operates to process information in an effective and non-trivial manner.


\section{Conclusion \label{sec:Conclusion}}

In summary, we have shown that the method of sequential parametric shifts used to turn well-stirred chemical systems into various logic gates~\cite{egbert_dynamic_2018} can be extended to non-well stirred systems where diffusion effects are important.  Using (predefined) spatiotemporal modulation of external conditions (e.g. feed rates), we explicitly show that it is possible to turn a uniform chemical medium into a full-adder.  More complicated gate networks are in principle possible.   Note also that getting the system to work at different scales (i.e. with wider gates) or with other parameters is likely possible, but would involve additional engineering, and would not contribute substantively to the primary contribution of the paper, which is to demonstrate a proof of concept, i.e. to show that the method we have presented (transforming a non-linear spatially distributed system into a network of computational units by modulating environmental dynamics) is possible. 

 A chemical implementation of the method presented here is an interesting possible future avenue of research.  The generality of the method makes it difficult to list the necessary properties a system must possess in order to apply it to real chemistry (as discussed in Ref.~\cite{egbert_dynamic_2018}), although some guidance may be gleaned from the model used in this paper.  Examples of sufficient (but not necessary) properties include a bistable regime (to implement binary logic), and other regimes (e.g. monostable) that can be accessed by tuning an external parameter (e.g. flow rate, light, etc).  Such external manipulation of a chemical system is certainly possible: for instance, NOR gates can be implemented using coupled micro-droplets and using light to control the oscillation state of the encapsulated Belousov-Zhabotinsky reaction inside the droplets~\cite{Wang_etal_2016} (note that the NOR gate implemented in Reg.~\cite{Wang_etal_2016} is not based on the method presented in this paper).   Another important aspect in the implementation of the method is the study of a uniform chemical medium subjected to two different contiguous external conditions.  Such an experiment would be a good test bed for the practicality of the method.


\appendix

\section{Mathematical analysis of information transfer between gates via diffusion}
\label{appendix:Info_transfer_diffusion}

\begin{figure}
  \centering
\includegraphics[width=0.49\textwidth]{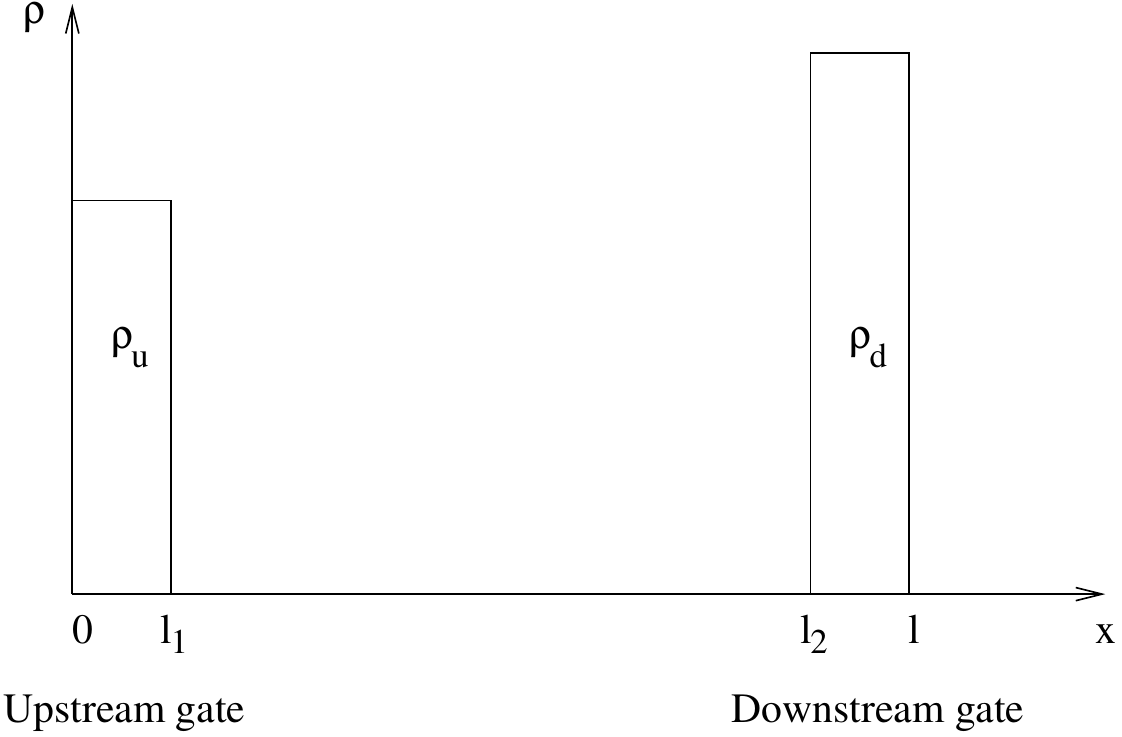}
  \caption{Gate setup.  After one computer clock tick, the upstream and downstream gates are in states $\rho_{u}$ and $\rho_{d}$, respectively.  The upstream gate then sends its output to the downstream gate via diffusion. \label{fig:Gate_setup}}
\end{figure}

Imagine two chemical gates that can communicate together via diffusion (see Fig.~\ref{fig:Gate_setup}). 
Let $\rho(x,t)$ be the space-time dependent concentration of a certain chemical.  In the following, we use high and low concentrations to determine the state of a gate.  A high concentration (say $\rho = 1$ in arbitrary units) represents a \true state while a low concentration (say $\rho = 0$) represents a \false state.  After one computer clock tick, the upstream and downstream gates are in states $\rho_{u}$ and $\rho_{d}$, respectively.  Before the next computer clock tick, the chemicals in the upstream gate diffuse toward the downstream gate and change its state.  For the computation to occur without error, the state of the downstream gate should be dictated solely by the chemicals diffusing toward it from the upstream gate and independent of its state at the end of a previous computation.

A relevant question to ask is, given the upstream and downstream gate states after one computer clock tick, is the state of the downstream gate more influenced by the input coming from the upstream gate or by its present state?  In other words, is the value $\rho(l_{2},t)$ more influenced by $\rho_{u}$ or $\rho_{d}$ after some finite time $t$? To answer this question, we solve the one-dimensional diffusion equation:
\begin{eqnarray}
\label{eq:Diffusion_equation}
\frac{\partial \rho(x,t)}{\partial t} & = & D\frac{\partial^{2}\rho(x,t)}{\partial x^{2}},
\end{eqnarray}
(where $D$ is a diffusion constant) subjected to the following boundary conditions:
\begin{eqnarray}
\label{eq:Boundary_conditions}
\rho(0,t) & = & \rho(l,t) \;\;=\;\;0 \;\;\;\;\;,\;\;\;\;\; t > 0
\end{eqnarray}
and initial condition:
\begin{equation}
\label{eq:Initial_condition}
\rho(x,0) = \rho_{u}\left(\theta(x) - \theta(x-l_{1})\right) + \rho_{d}\left(\theta(x-l_{2}) - \theta(x - l)\right), 
\end{equation}
where $\rho_{u}$ and $\rho_{d}$ are constants representing the state of the upstream and downstream gates.  The solution can be obtained using the method of separation of variables (e.g. \cite{Nagle_Saff_2017}):
\begin{eqnarray}
\rho(x,t) & = & 
\sum_{m=1}^{\infty} \frac{2}{m\pi}\bigg[\rho_{u}\left(1 - \cos\left(\frac{m\pi l_{1}}{l}\right)\right) \nonumber \\
& & + \rho_{d}\left((-1)^{m+1} + \cos\left(\frac{m\pi l_{2}}{l}\right)\right)\bigg] \nonumber \\
&   & \times \sin\left(\frac{m\pi}{l}x\right) e^{-\left(\frac{m^{2}\pi^{2}}{l^{2}}\right)Dt}.
\end{eqnarray}
Note that the dependence in time in the above solution goes as $e^{-m^{2}t}$.  For $t>0$, we see that terms with larger values of $m$ are suppressed more strongly than those with lower values of $m$.  We thus conclude that the $m = 1$ term probably dictates the bulk of the dynamics at finite times.  Specializing to the spatially symmetric gate configuration $l_{2} = l - l_{1}$, the value of the concentration at the downstream gate is given by:
\begin{eqnarray}
\rho(l_{2},t) & = &  \sum_{m=1}^{\infty} \frac{2}{m\pi}\bigg[\bigg(\rho_{u} + (-1)^{m+1}\rho_{d}\bigg) \nonumber \\
& & \left(1 - \cos\left(\frac{m\pi l_{1}}{l}\right)\right)\bigg]  
\sin\left(\frac{m\pi}{l}l_{2}\right) e^{-\left(\frac{m^{2}\pi^{2}}{l^{2}}\right)Dt}. \nonumber \\
\end{eqnarray}
Since the bulk of the dynamics is dictated by the $m=1$ term, we have:
\begin{eqnarray}
\rho(l_{2},t) & \propto & (\rho_{u} + \rho_{d}).
\end{eqnarray}
The above implies that it is not possible to distinguish the following cases: ($\rho_{u} = 1$, $\rho_{d}=0$) and ($\rho_{u} = 0$, $\rho_{d} = 1$).  Thus pure diffusion does not seem to be sufficient to transmit information between gates in a chemical computer when multiple clock ticks are required.

\section{Details of the NAND operation}
\label{appendix:Details_NAND_operation}

We now describe in detail the entire regime sequence of a NAND gate, starting with receipt of input from two upstream gates. For simplicity, we assume that these upstream gates are also NAND gates but, as we shall see later, we can relax this assumption.

Our explanation starts with the gate in the `receive' step, where it is placed in regime RC (monostable) for 1 time unit (see Table~\ref{tab:NAND_regime_sequence_specification}). Its neighbouring upstream gates are shifted by $9.6$ time units and so are currently in the eighth NAND-gate regime (`transmit'), where it is in regime RA (bistable) (see Table~\ref{tab:NAND_regime_sequence_specification}).

Figure~\ref{fig:cross_section_NAND} in the main text shows the evolution of a one-dimensional reaction-diffusion-based NAND gate. Each column in this figure shows the `downstream' gate flanked by two `upstream' gates that provide the input to the downstream NAND gate. The upstream gates are initialized to either \UVtrue or \UVfalse according to the four possible inputs indicated at the bottom of each column, and again, the gates follow an identical periodic sequence of feed rate configurations (see Table~\ref{tab:NAND_regime_sequence_specification} in the main text) but, as just described, the regime sequences are shifted by $9.6$ time units.

At $t=0$, the downstream gate is in the monostable configuration and so if it were isolated, its state would approach $(U,V)=0,0$. However, the gate is not isolated and chemicals from the upstream gates diffuse into the downstream gate, causing the middle of the gate to come to a steady-state that is approximately equivalent to a scaled linear combination of the two input gates steady states~\cite{Note1}.


The first step in our implementation of the NAND operation (see Ref.~\cite{egbert_dynamic_2018}) is the \textbf{categorization} of the input, so that the initial conditions that are associated with an output of \false are at one fixed point, and the other initial condition goes to the other. In the non-spatial system, this was accomplished by a threshold dynamic, where $V$ above a certain threshold value (\ie above the separatrix in the bistable system) approaches the \UVtrue, while all other values of $V$ go to \UVfalse. In the non-spatial system, this nicely separates the input initial conditions by their output class, causing the initial condition associated with an input of F/F to go to \UVfalse and the other three inputs to go to \UVtrue. Unfortunately, this does not work in the spatial system. Because the information transmission is a diffusion process, whenever the upstream gate is providing an input of \true, the edges of the downstream gate (see \eg the area marked by an `e' in Figure~\ref{fig:cross_section_NAND} in the main text) will have higher concentration of $V$ than the center of the gate (marked `c'). The center of the T/T downstream gate has a higher concentration of $V$ than the center of any of the other downstream gates, but the concentration of $V$ at the edge of this gate, and problematically also in the T/F and F/T gates will be higher still. There is therefore, at this stage, no simple threshold value of $V$ that allows us to categorize the inputs by their target output, \ie to distinguish input T/T from inputs F/F, T/F, and F/T the way that there was in the non-spatial system.

To resolve this, we place the downstream gate into the ``slow monostable'' configuration and the upstream gates into the monostable configuration. The combination of these two regimes, causes the chemical concentrations near the edges of the downstream gate to approach $(U,V)=0$ more rapidly than those at its center ($t
\in 1 \rightarrow 1.4$), and although the values of $V$ at the edge are still higher than in the center at the end of this phase, they continue to drop at the start of the next ``categorize'' phase, and the gates evolve such that only the gate with T/T input contains states that are above the separatrix, and the inputs are correctly categorized ($t \in 1.4 \rightarrow 6.4$).

Note that the effective operation of this first phase of our system depends upon the gates having the correct width relative to the rate at which the reactants diffuse. We simulated full-adder networks (described below) using various gate widths in the range $\{1.9,2.0 ... 3.4, 3.5\}$ length units.  Those constructed with widths between $2.0$ and $3.4$ (inclusive) operated correctly, but those constructed with gate widths of $1.9$ or $3.5$ failed. It may be the case that the range of functional operating widths could be extended by changing the duration of certain regimes, but we have not conducted an investigation into this.

The second step in our implementation of the NAND operation is the \textbf{inversion} of the categorized inputs, where the trajectories that are near \UVfalse move to \UVtrue and vice versa. We accomplish this in the same way as in the non-spatial system described in Ref.~\cite{egbert_dynamic_2018}. First, a brief transient in the perturbatory configuration, increases $V$, bumping the F/F, T/F and F/T systems out of the \UVfalse basin of attraction. A subsequent transient in the bistable regime (``invert'') causes the concentration of $V$ in the T/T trajectory to move above the concentration of $V$ in the others (see $t=6.75$ in Fig.~\ref{fig:cross_section_NAND} in the main text and also frames B and C in Fig.~4 of Ref.~\cite{egbert_dynamic_2018}). The system is then placed briefly in the monostable regime, which causes all of the trajectories to decrease in $V$. The timing of this transient is selected so that at the end of the transient, when the system enters the bistable regime (``stabilize''), the higher-in-V trajectories (those corresponding to inputs of F/F, T/F and F/T) are above the separatrix and the other trajectory (input T/T) is below it. This final phase causes the system to fall into the correct NAND output states where the trajectories started with inputs F/F, T/F and F/T approach \UVtrue and the trajectory started with the T/T input approaches \UVfalse.  At this point, the downstream gate is now ready to act as the upstream gate for whatever gates are to accept its input.  A video showing the operation of a NAND gate (composed of a ``receiving'' gate flanked by two ``transmitting'' gates) can be seen in the Supplementary Information (Movie S2).

\section{Other logic gates}
\label{appendix:Other_gates}
NOR was produced by removing the ``shrink-boundaries'' step. Removal of this regime causes any downstream node with one or more \true inputs to be above threshold. AND and OR gates were produced by removing the inversion portion of the NAND and NOR gates respectively. For all of these modifications, the ``stabilize'' regime was extended by the amount necessary such that the total duration for the computation phase of each gate-type remained the same.

\section*{Competing interests \label{sec:Competing_interests}}

The authors have no competing interests.

\section*{Authors' contribution \label{sec:Contribution}}

ME conceived of the study, carried out the numerical simulations, and participated in the writing of the manuscript; JSG conceived of the study, contributed in the more theoretical aspects of the study, and participated in the writing of the manuscript; JPM coordinated the study, participated in its design, and helped draft the manuscript.  All authors gave final approval for publication.

\begin{acknowledgments}
The authors thank J. Szymanski for useful discussions.
\end{acknowledgments}

\section*{Funding \label{sec:Funding}}

This research is supported by Repsol S.A.


\end{document}